\documentclass[aps,twocolumn,groupedaddress]{revtex4}
\usepackage{epsfig}

\newcommand{\im}{\textrm{Im}}

\newcommand{\D}[1]{\; d #1 \;}

\newcommand{\lbl}[1]{\label{eq:#1}}
\newcommand{ \rf}[1]{(\ref{eq:#1})}

\newcommand{\be}{\begin{equation}}
\newcommand{\ee}{\end{equation}}
\newcommand{\bea}{\begin{eqnarray}}
\newcommand{\eea}{\end{eqnarray}}
\newcommand{\setl}{\setlength\arraycolsep{2pt}}

\newcommand{\noi}{\noindent}
\newcommand{\nn}{\nonumber}
\newcommand{\ra}{\rightarrow}

\newcommand{\lesssim}{ {\
\lower-1.2pt\vbox{\hbox{\rlap{$<$}\lower5pt\vbox{\hbox{$\sim$}}}}\ } 
}
\newcommand{\gtrsim}{ {\
\lower-1.2pt\vbox{\hbox{\rlap{$>$}\lower5pt\vbox{\hbox{$\sim$}}}}\ } 
}

\newcommand{\cM}{{\cal M}}

\newcommand{\cO}{{\cal O}}

\newcommand{\Imm}{\mbox{\rm Im}}

\newcommand{\MeV}{\mbox{\rm MeV}}
\newcommand{\GeV}{\mbox{\rm GeV}}

\newcommand{\annd}{\mbox{\rm and}}

\newcommand{\als}{\alpha_{\mbox{\rm {\scriptsize s}}}}

\newcommand{\exxp}{\mbox{\rm \tiny exp.}}

\newcommand{\NQCD}{QCD$_{\infty}~$}
\newcommand{\thh}{\mbox{\rm {\tiny th}}}

\input epsf

\begin{document}
\bibliographystyle{revtex}

\preprint{CPT-2000/P.4047}
\preprint{UAB-FT-494}



\title{Tests of Large--$N_c$ QCD from Hadronic $\tau$ Decay}



\author{S. Peris$~^a$, B. Phily$~^b$ and E. de Rafael$~^b$}
\affiliation{
$^a$ Grup de F{\'\i}sica Te{\`o}rica and IFAE, Universitat Aut{\`o}noma
de Barcelona, E-08193 Barcelona, Spain\\
$^b$Centre de Physique Th{\'e}orique, CNRS--Luminy, Case 907,
F-13288 Marseille Cedex 9, France}



\begin{abstract} We use the Aleph data on vector and axial-vector
  spectral functions to test simple duality properties of  QCD in the
large $N_c$ limit, which emerge in the approximation of a {\it minimal
hadronic ansatz} of a spectrum of narrow states. These duality properties
relate the short--  and long--distance behaviours of specific correlation
functions, which are order parameters of spontaneous chiral symmetry
breaking, in a  way that we find well supported by the data.  
\end{abstract}
\pacs{11.15Pg, 12.38Aw, 12.20Fv}

\maketitle

{\bf 1.}~At first sight, the {\it hadronic world} predicted by QCD in the
limit of a large number of colours $N_c$~\cite{tH74} may seem rather
different from the real world. The hadronic spectrum of vector and
axial--vector states, observed e.g. in $e^{+} e^{-}$ annihilations and in
$\tau$ decays, has certainly much more structure than the infinite set of
narrow states predicted by large $N_c$ QCD~\cite{Wi79} (\NQCD). There are,
however, many instances in Particle Physics where one is only
interested in certain weighted integrals of hadronic spectral
functions. In these cases, it may be enough to know a few {\it global}
properties of the hadronic spectrum; one does not expect the integrals
to depend crucially on the details of
the spectrum at all energies. Typical examples of that are the coupling
constants of the effective chiral Lagrangian of
QCD at low energies, as well as the coupling constants of
the effective chiral Lagrangian of the electroweak interactions of
pseudoscalar particles in the Standard Model, which are needed to
understand
$K$--Physics in particular, (see e.g. the review article in
ref.~\cite{Pi99} and references therein.) It is in these examples that the
{\it hadronic world} predicted by \NQCD may provide a good approximation to
the real hadronic spectrum. If so,
\NQCD could then become a useful phenomenological approach for understanding
non--perturbative QCD physics at low energies.

There are indeed a number of successful
calculations which have already been made
within the framework of \NQCD, (see ref.~\cite{KPdeR00} and references
therein.) The picture which emerges from these applications is one of a
remarkable simplicity. It is found that, when dealing with Green's
functions that are {\it order parameters} of spontaneous chiral symmetry
breaking, the restriction of the infinite set of large $N_c$ narrow states
to a {\it minimal hadronic ansatz} which is needed to satisfy the leading
short-- and long--distance behaviours of the relevant Green's functions,
provides already a very good approximation to the observables one
computes. The purpose of this note is to investigate this
{\it minimal hadronic ansatz} approximation in a case
where one can compare, in detail, the theoretical predictions to
the phenomenological results evaluated with experimental
data.

{\bf 2.}~Of particular interest for our purposes is the correlation 
function ($Q^2\equiv -q^2\ge 0$ for $q^2$ space--like)
\be\lbl{lrtpf}
\Pi_{LR}^{\mu\nu}(q)= 
 2i\int d^4 x\,e^{iq\cdot x}\langle 0\vert
\mbox{\rm T}\left(L^{\mu}(x)R^{\nu}(0)^{\dagger}
\right)\vert 0\rangle\,,
\ee 
with colour singlet currents
\be
R^{\mu}\left(L^{\mu}\right)=
\bar{d}(x)\gamma^{\mu}\frac{1}{2}(1\pm\gamma_{5})u(x)\,.
\ee 
In the chiral limit, $m_{u,d,s}\ra 0$\, , this correlation function has
only a transverse component, 
\be\lbl{lritpf}
\Pi_{LR}^{\mu\nu}(Q^2)=(q^{\mu}q^{\nu}-g^{\mu\nu}q^2)\Pi_{LR}(Q^2)\,.
\ee
The self-energy--like function $\Pi_{LR}(Q^2)$  vanishes order by order
in  perturbative QCD (pQCD) and
is an order parameter of S$\chi$SB for all values of $Q^2$;
therefore it obeys an unsubtracted dispersion relation,
\be\lbl{disprel}
\Pi_{LR}(Q^2)=\int_{0}^{\infty}dt\frac{1}{t+Q^2}
\frac{1}{\pi}\Imm\Pi_{LR}(t)\,.
\ee

In \NQCD the spectral function $\frac{1}{\pi}\Imm\Pi_{LR}(t)$ consists 
of
the difference of an infinite number of narrow vector and axial--vector
states, together with the Goldstone pole of the pion:
\bea
\lefteqn{\frac{1}{\pi}\Imm\Pi_{LR}(t) =   
\sum_{V}f_{V}^2 M_{V}^2\delta(t-M_{V}^2)} \nn \\
 & & -F_{0}^2\delta(t) -\sum_{A}f_{A}^2
M_{A}^2\delta(t-M_{A}^2)\,.
\eea 
The low $Q^2$ behaviour of $\Pi_{LR}(Q^2)$, i.e.
the long--distance behaviour of the correlation function in
Eq.~\rf{lrtpf}, is governed by chiral perturbation theory:
\be 
-Q^2\Pi_{LR}(Q^2)\vert_{Q^2\ra 0}=F_{0}^2+4L_{10}Q^2+\cO(Q^4)\,,
\ee 
where $F_{0}$ is the pion coupling constant in the chiral limit, and
$L_{10}$ is one of the coupling constants
of the $\cO(p^4)$ effective chiral Lagrangian. 
The high $Q^2$
behaviour of $\Pi_{LR}(Q^2)$, i.e.
the short--distance behaviour of the correlation function in
Eq.~\rf{lrtpf}, is governed by the operator product expansion (OPE) 
of the
two local currents in Eq.~\rf{lrtpf}~\cite{SVZ79},
\be\lbl{OPE}
\lim_{Q^2\ra\infty}Q^6\Pi_{LR}(Q^2)= 
 \left[-4\pi^2\frac{\alpha_s}{\pi}+\cO(\alpha_s^2)\right]
\langle\bar{\psi}\psi\rangle^2\,,
\ee
which implies the two Weinberg sum rules:
\be\lbl{weinbergsr1}
\int_{0}^{\infty}\!\!dt \, \Imm\Pi_{LR}(t)=\sum_{V}f_{V}^2
M_{V}^2-\sum_{A}f_{A}^2 M_{A}^2-\!F_{0}^2=0\,,
\ee
and
\be\lbl{weinbergsr2}
\int_{0}^{\infty}\!\!dt \, t\, \Imm\Pi_{LR}(t)=\sum_{V}f_{V}^2
M_{V}^4-\sum_{A}f_{A}^2 M_{A}^4=0\,,
\ee 
as well as the sum rule~\cite{KdeR98}
\be\lbl{KdeRsr}
\sum_{V}\! f_{V}^2 M_{V}^6 -\! \sum_{A}\! f_{A}^2
M_{A}^6= \left[-4\pi\alpha_s +\cO(\alpha_s^2)\right]
\langle\bar{\psi}\psi\rangle^2\,.
\ee

In fact, as pointed out in ref.~\cite{KdeR98}, in \NQCD there exist an
infinite number of Weinberg--like sum rules. In full
generality, the moments of the
$\Pi_{LR}$ spectral function with $n=3,4,\dots$, 
\bea\lbl{posmoments}
\lefteqn{\!\!\!\!\!\!\!\!\!\!\!\!\!\!\!\!\!\!
\int_{0}^{\infty} dt\,t^{n-1}\left[\frac{1}{\pi}\Imm\Pi_{V}(t)-
\frac{1}{\pi}\Imm\Pi_{A}(t)\right]=} \nn \\
 & & \sum_{V} f_{V}^2 M_{V}^{2n} - \sum_{A} f_{A}^2 M_{A}^{2n}\,,
\eea
govern the short--distance expansion of the $\Pi_{LR}(Q^2)$
function
\bea\lbl{largeQ}
\lefteqn{\!\!\!\!\!\!\!\!\!
\Pi_{LR}(Q^2)\vert_{Q^2\ra\infty} = 
\left(\sum_{V} f_{V}^2 M_{V}^6 - \sum_{A} f_{A}^2
M_{A}^6\right)\frac{1}{Q^6}} \nn
\\ &  & + \left(\sum_{V} f_{V}^2 M_{V}^8 - \sum_{A} f_{A}^2
M_{A}^8\right)\frac{1}{Q^8} +
\cdots \,.
\eea
On the other hand inverse moments of the
$\Pi_{LR}$ spectral function with the pion pole removed (which we denote
by $\Imm\tilde{\Pi}_{A}(t)$) determine a class of  
coupling constants
of the low--energy effective chiral Lagrangian. For example,
\bea\lbl{invmom}
\lefteqn{\!\!\!\!\!\!\!\!\!
\int_{0}^{\infty} dt \frac{1}{t}\left[\frac{1}{\pi}\Imm\Pi_{V}(t)-
\frac{1}{\pi}\Imm\tilde{\Pi}_{A}(t)\right]=} \nn \\
 & & \sum_{V} f_{V}^2 -\sum_{A} f_{A}^2 = -4 L_{10}\,.
\eea
Moments with higher inverse powers of $t$ are associated with 
couplings of
composite operators of higher dimension in the chiral Lagrangian. Tests of
the two Weinberg sum rules in Eqs.~\rf{weinbergsr1} and \rf{weinbergsr2}
and of the
$L_{10}$ sum rule in Eq.~\rf{invmom}, in a different context from the one we
are interested in here, have often appeared in the literature (see e.g.
refs.~\cite{DHGS98} and \cite{DS99} for recent discussions where earlier
references can also be found).

{\bf 3.}~The {\it minimal hadronic ansatz} which satisfies the two
Weinberg sum rules in Eqs.~\rf{weinbergsr1} and \rf{weinbergsr2} is a
spectrum of one vector state $V$, one axial--vector state $A$ and the
Goldstone pion, with the ordering~\cite{KdeR98} $M_{V}< M_{A}$.  
In this approximation,
$\Pi_{LR}(Q^2)$ has a very simple form
{\setl
\bea 
\!\!\!\!\!\!\!\!\!\!\!\!\!\!\!\!-Q^2
\Pi_{LR}(Q^2) & = & 
\frac{F_{0}^2}{\left(1+\frac{Q^2}{M_{V}^2}\right)
\left(1+\frac{Q^2}{M_{A}^2}\right)}  \\
  &  = &  \frac{M_{A}^2
M_{V}^2}{Q^4}\frac{F_{0}^2}{\left(1+\frac{M_{V}^2}{Q^2}\right)
\left(1+\frac{M_{A}^2}{Q^2}\right)}\,.
\eea}

\noi 
This equation shows, explicitly, a remarkable short--distance
$\rightleftharpoons$ long--distance duality~\cite{deR99}. Indeed, 
with $g_{A}$ defined
so that $M_{V}^2=g_{A}M_{A}^2$ and
$z\equiv\frac{Q^2}{M_{V}^2}$,  the non--local order
parameters corresponding to the long--distance expansion for $z\ra 0$,
which are couplings of the effective chiral Lagrangian i.e., 
\bea\lbl{chiralex}
\lefteqn{ 
-Q^2\Pi_{LR}(Q^2)\vert_{z\ra 0}=
F_{0}^2\left\{1-(1+g_{A})z\right.} \nn \\
 & & \left. +(1+g_{A}+g_{A}^2)z^2+\cdots\right\}\,,
\eea 
are correlated to the  local order parameters of the
short--distance OPE for $z\ra\infty$ in a very simple
way:
\bea\lbl{opemha}
\lefteqn{\!\!\!\!\!\!\!\!\!\!\!\!\!\!\!\!\!\!\!\!\!-Q^2
\Pi_{LR}(Q^2)\vert_{z\ra\infty}=
F_{0}^2\frac{1}{g_{A}}\frac{1}{z^2}
\left\{1- \left(1+\frac{1}{g_{A}}\right)\frac{1}{z}\right.} \nn \\
& & \left. +
\left(1+\frac{1}{g_{A}}+\frac{1}{g_{A}^2}\right)\frac{1}{z^2}
+\cdots\right\}\,;
\eea
in other words, there is a one-to-one correspondance 
between the two expansions
by changing
\be
g_{A}\rightleftharpoons \frac{1}{g_{A}}\quad \annd \quad
z^{n}\rightleftharpoons \frac{1}{g_{A}}\frac{1}{z^{n+2}}\,.
\ee
The moments of the $\Pi_{LR}$ spectral function, when evaluated in
the  {\it minimal hadronic ansatz} approximation, can be converted
into a very simple set of finite energy sum rules (FESR's), corresponding
to the OPE in Eq.~\rf{opemha}  
{\setl
\bea\lbl{moments2}
\int_{0}^{s_0}
dt\,t^{2}\frac{1}{\pi}\Imm\Pi_{LR}(t) & =
&-F_{0}^{2}M_{V}^4\frac{1}{g_{A}}\,,
\\ \lbl{moments3}
\int_{0}^{s_0}
dt\,t^{3}\frac{1}{\pi}\Imm\Pi_{LR}(t) & =
&-F_{0}^{2}M_{V}^6\frac{1\!+\!\frac{1}{g_{A}}}{g_{A}}\,, \\
\lbl{moments4}
\int_{0}^{s_0}
dt\,t^{4}\frac{1}{\pi}\Imm\Pi_{LR}(t) & =
&-F_{0}^{2}M_{V}^8\frac{1\!+\!\frac{1}{g_{A}}\!+\!\frac{1}{g_{A}^2}}{g_{A}}\,,
\\
\cdots\qquad\qquad &  &\qquad\cdots\,\,. \nn
\eea}

\vspace{-0.5cm}
\noi
where the upper limit of integration $s_{0}$ denotes the onset
of the pQCD continuum which, in the chiral limit, is common to the
vector and axial--vector spectral functions. It is important to realize
that $s_{0}$ is not a free parameter. Its value is fixed by the requirement
that the OPE of the correlation function of two vector currents, (or two
axial--vector currents,) in the chiral limit, have no
$1/Q^2$ term, which results in an implicit equation for
$s_{0}$~\cite{BLdeR85,PPdeR98}. In the
{\it minimal hadronic ansatz} approximation the onset of the pQCD continuum,
which we shall call
$s_{0}^{*}$, is then fixed by the equation
\be\lbl{duality}
\frac{N_c}{16\pi^2}\frac{2}{3}s_{0}^{*}\left(1+\cO(\als)\right)=F_{0}^2
\frac{1}{1-g_{A}}\,.
\ee
Also, the moments which
correspond to the chiral expansion in Eq.~\rf{chiralex} are given by
another simple set of FESR's: {\setl
\bea\lbl{inversemoments}
\int_{0}^{s_0}
dt\,\frac{1}{\pi}\Imm\tilde{\Pi}_{LR}(t) & =
& F_{0}^2\,,
\\
\lbl{inversemoments2}
\int_{0}^{s_0}
\frac{dt}{t}\,\frac{1}{\pi}\Imm\tilde{\Pi}_{LR}(t) & =
& \frac{F_{0}^2}{M_{V}^2}(1\!+\!g_{A})\,, \\
\lbl{inversemoments3}
\int_{0}^{s_0}
\frac{dt}{t^2}\,\frac{1}{\pi}\Imm\tilde{\Pi}_{LR}(t) & =
& \frac{F_{0}^2}{M_{V}^4}(1\!+\!g_{A}\!+\!g_{A}^2)\,,
\\
\cdots\qquad\qquad &  &\qquad\cdots\,. \nn
\eea}

\vspace{-0.5cm}
\noi  
We propose to test these duality relations by comparing  moments of
the \underline{physical} spectral function
$\frac{1}{\pi}\Imm\Pi_{LR}(t)$ to the predictions of the {\it minimal
hadronic ansatz}.

{\bf 4.}~The ALEPH collaboration at LEP has measured the
inclusive invariant mass--squared distribution of hadronic $\tau$
decays~\cite{ALEPH} into non--strange particles. They have been able to
extract from their data both the vector current spectral function
$\frac{1}{\pi}\Imm\Pi_{V}^{\exxp}(t)$ and the axial--vector current
spectral function
$\frac{1}{\pi}\Imm\Pi_{A}^{\exxp}(t)$ up to 
$t\simeq 3\,\GeV^2$. In fact, in the real world, the correlation
function in Eq.~\rf{lritpf} has a non--transverse term as well,
which is dominated by the pion pole contribution to the axial--vector
component. The vector contribution to the non--transverse term vanishes in
the limit of isospin invariance.

In order to compare the moments of the experimental  spectral
function
$\frac{1}{\pi}\Imm\Pi_{LR}^{\exxp}(t)$ to those in
Eqs.~\rf{moments2}-\rf{moments4} and 
\rf{inversemoments}-\rf{inversemoments3} 
we still have to correct for the fact that the
FESR's in these equations apply in the chiral limit where
$m_{u,d}\ra 0$. This we do by exploiting the analyticity properties of
the two--point function $\Pi_{LR}$ in the complex $q^2$--plane.
Integration over a standard contour relates weighted integrals of the
spectral function $\frac{1}{\pi}\Imm\Pi_{LR}^{\exxp}(t)$ in a finite
interval on the real axis to integrals of
$\Pi_{LR}(q^2)$ over a {\it small} circle
$\vert q^2\vert=s_{\thh}$ and a {\it large} circle $\vert q^2\vert=s_{0}$:
\bea\lbl{contour}
\int_{s_{th}}^{s_0} \D{t} f(t) \im \Pi_{LR} (t)  = \qquad\qquad\nn \\ 
\frac{1}{2i}\!\!\!\!\oint\limits_{|q^2|=s_{th}} \!\!\!\!\!\!dq^2 f(q^2)
\Pi_{LR} (q^2)\!-\!\!\frac{1}{2i}\!\!\!\!\oint\limits_{|q^2|=s_0} 
\!\!\!\!\!\!dq^2 f(q^2)
\Pi_{LR} (q^2)\,,
\eea

\noi
where the weight function $f(q^2)$ is a conveniently chosen analytic
function inside the contour; in our case simple powers and inverse powers
of $q^2$. The chiral corrections in the {\it small} circle are particularly
important in the evaluation of the inverse moments. We have evaluated them
by taking into account the one loop expression of
$\Pi_{LR}(z)$ in chiral perturbation theory~\cite{GL84}. The chiral
corrections in the {\it large} circle are rather small. They appear as
leading $1/Q^2$ and next--to--leading
$1/Q^4$ power corrections in the OPE of $\Pi_{LR}(Q^2)$ at large
$Q^2$ but their coefficients, proportional to quark masses, are
small~\cite{FNdeR79}. With these corrections incorporated, we proceed
now to the comparison we are looking for. This is shown in Figs.~1 and 2
below where we show the various moments as a function of $s_{0}$.

\vspace{0.5cm}
\centerline{\epsfbox{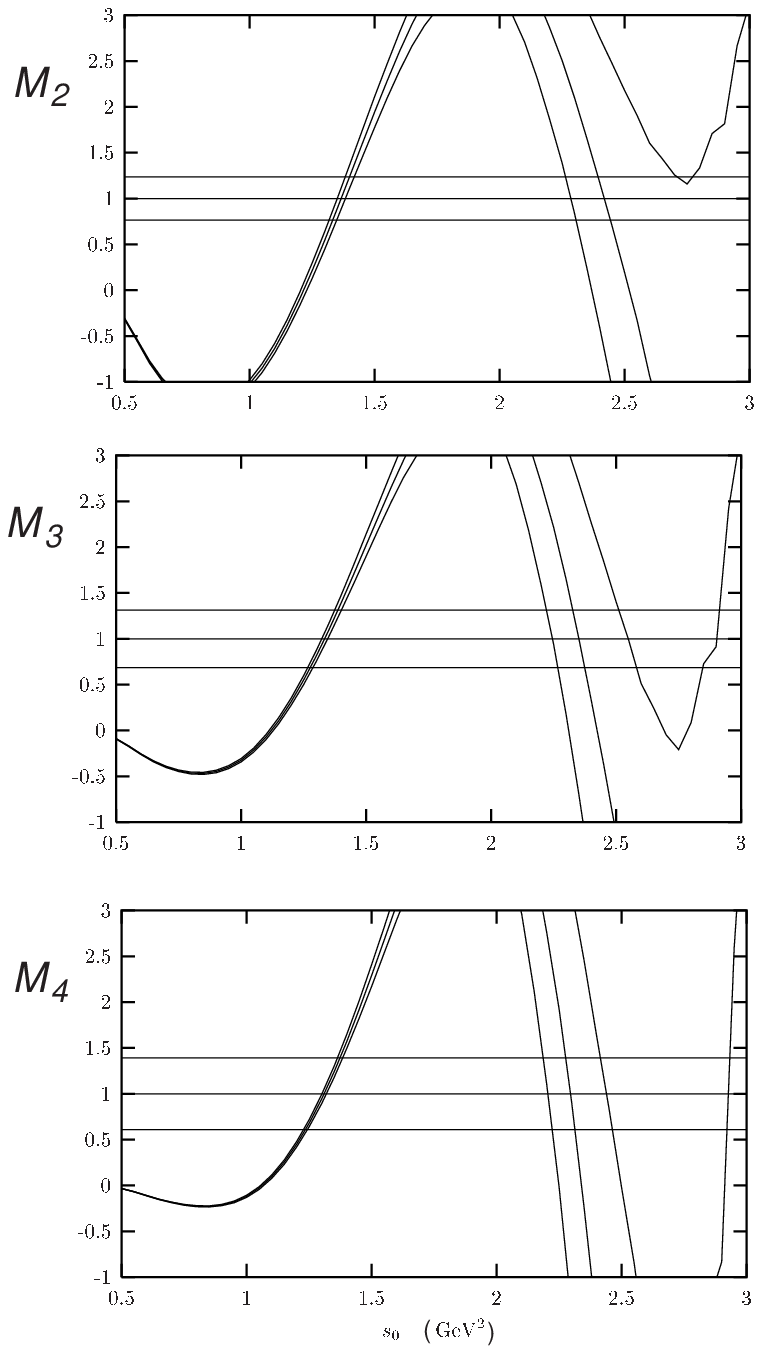}}
\vspace{0.3cm}
{\bf{Fig.~1}} {\it Plot of the experimental moments in Eqs.~\rf{moments2}, 
\rf{moments3} and \rf{moments4} normalized to the 
{\it minimal hadronic ansatz}
predictions on the r.h.s.}
\vspace{0.5cm}

\noi
The six plots in Figs.~1 and 2 show the experimental
moments on the l.h.s. of Eqs.~\rf{moments2}-\rf{moments4} and 
Eqs.~\rf{inversemoments}-\rf{inversemoments3}, respectively,  
as a function of $s_{0}$, extrapolated to the chiral limit as
discussed above and normalized to the corresponding {\it minimal hadronic
ansatz} predictions on the r.h.s. 


The horizontal bands on
these plots correspond to the induced error of the {\it minimal hadronic
ansatz} predictions
from the input values:
$F_{0}=87\pm3.5\,\MeV$, $M_{V}=748\pm29\,\MeV$ and $g_{A}=0.50\pm 0.06$.
These are the values favored by a global fit of the {\it minimal hadronic
ansatz} to low--energy observables~\cite{PPdeR98}. The moments
$\cM_n$, with the experimental error propagation included, are the curved
bands in the figures.

\vspace{0.5cm}
\centerline{\epsfbox{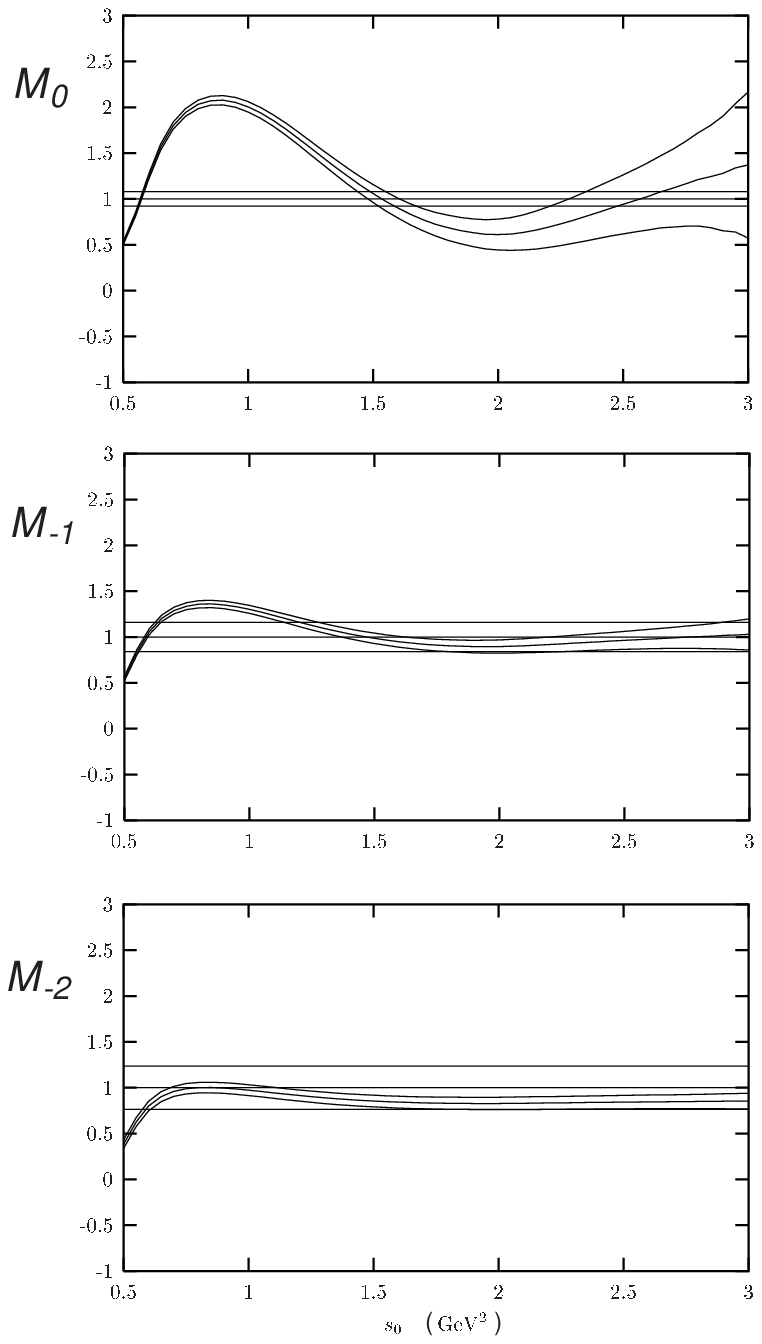}}
\vspace{0.3cm}
{\bf{Fig.~2}} {\it Plot of the experimental moments in Eqs.~\rf{inversemoments},
\rf{inversemoments2} and  \rf{inversemoments3} normalized to the 
{\it minimal hadronic ansatz}
predictions on the r.h.s.}
\vspace{0.5cm}

The remarkable feature which the curves in Figs.~1 and 2 show is that,
within errors, the first crossing of \underline{all} the experimental
moments with the {\it minimal hadronic ansatz} band takes place in the
\underline{same}
$s_{0}$ region, around $s_{0}\sim 1.4~\GeV^2$ rather close indeed to the
$s_{0}^{*}$ value which follows from Eq.~\rf{duality}:
$s_{0}^{*}\!\!=(1.2\pm 0.2)~\GeV^2$. We have also checked that for the 2nd
Weinberg sum rule in Eq.~\rf{weinbergsr2}, not shown in the figures. In
fact, the agreement for the inverse moments is excellent. This is due to
the fact that inverse moments put more and more weight on the low energy
tail of the spectral function, which is known to be dominated by the
$\rho$--resonance. By contrast, the
positive moments are very sensitive to the cancellations between opposite
parity hadronic states; this is why the experimental curves show larger
and larger oscillations as one increases the power of the moment.
In spite
of that, it is quite impressive that, when restricted to the $s_{0}$ region
of duality, the experimental moments agree well with the {\it minimal
hadronic ansatz} prediction, even for rather large powers which correspond
to vacuum expectation values of operators of higher and higher dimension
in the OPE. 

We conclude that the experimental data from ALEPH is consistent with the
simple pattern of duality properties between short and
long--distances which follow from the  {\it minimal hadronic ansatz} of a
narrow vector and axial-vector states plus the Goldstone
pion in large--$N_c$ QCD.


\vskip 0.1cm

Work supported in part by TMR, EC-Contract No. ERBFMRX-CT980169 
(EuroDa$\phi$ne). The work of
S.P. is also supported by CICYT-AEN99-0766.
\bibliography{references}

%
%

%
%

\end{document}